\documentclass{article}

\usepackage[final]{neurips_2020}

\usepackage{natbib}
\usepackage[utf8]{inputenc} 
\usepackage[T1]{fontenc}    
\usepackage{hyperref}       
\usepackage{url}            
\usepackage{booktabs}       
\usepackage{amsfonts}       
\usepackage{nicefrac}       
\usepackage{microtype}      
\title{Nose to Glass: Looking In to Get Beyond}

\author{%
  Josephine Seah\\
  Centre for AI and Data Governance\\
  School of Law\\
  Singapore Management University\\
  \texttt{shseah@smu.edu.sg} \\
}

\begin{document}

\maketitle

\begin{abstract}
Brought into the public discourse through investigative work by journalists and scholars, awareness of algorithmic harms is at an all-time high. An increasing amount of research has been dedicated to enhancing responsible artificial intelligence (AI), with the goal of addressing, alleviating, and eventually mitigating the harms brought on by the roll out of algorithmic systems. Nonetheless, implementation of such tools remains low. Given this gap, this paper offers a modest proposal: that the field--particularly researchers concerned with responsible research and innovation--may stand to gain from supporting and prioritising more ethnographic work. This embedded work can flesh out implementation frictions and reveal organisational and institutional norms that existing work on responsible artificial intelligence has not yet been able to offer. In turn, this can contribute to more insights about the anticipation of risks and mitigation of harm. This paper reviews similar empirical work typically found elsewhere--commonly in science and technology studies and safety science research--and lays out challenges of this form of inquiry.   
\end{abstract}

\section{Introduction}

Awareness of algorithmic harms is at an all-time high. Response to concerns over these harms initially led to widespread publication of ethical principles in recent years \citep{jobinGlobalLandscapeAI2019}. These efforts have nonetheless been criticised as ill-defined and non-binding \citep{yeung2019ai, ochigameInventionEthicalAI2019, mittelstadtAIEthicsToo2019}. Nonetheless, it is clear that there has been a shift towards operationalising these principles, along with an ever-increasing number of tools and methods aimed at realising the goal of responsible artificial intelligence (AI) development and use \citep{morleyWhatHowInitial2019,brundageTrustworthyAIDevelopment2020a}. These tools include technical work done under the umbrella of Explainable AI (XAI) \citep{dasOpportunitiesChallengesExplainable2020} or interpretability techniques, such as “glassbox” machine learning (ML) models \citep{jungSimpleRulesComplex2017} or post-hoc explanation techniques \citep{lundbergUnifiedApproachInterpreting2017}. Processual innovations have also been suggested, most notably by scholars calling for the use of documentation processes to encourage more transparency in the production pipeline \citep{gebruDatasheetsDatasets2020, mitchellModelCardsModel2019, rajiMLAnnotationBenchmarking2020}. At the same time, similar concerns have led to efforts within the AI Research community to take steps towards more responsible and ethical research: be it through the the convening of conferences like the ACM Conference on Fairness, Accountability and Transparency or the AAAI/ACM Conference on Artificial Intelligence, Ethics, and Society; or through changes to conference requirements that ask researchers to think more intentionally about the societal implications of their work.   

Scholars have also been calling for more forms of multidisciplinary research. This call has been split between the urgency for social scientists’ increased involvements to addressing the alignment question--such as \citet{irvingAISafetyNeeds2019}'s call for psychologists to be working in the field; and the urgency for social scientists to contribute towards ensuring that the gains of data science can be directed towards socially just goals \citep{dignazioDataFeminism2020a}. This paper focuses on the latter: more specifically, taking into account \citet[p.330]{sloaneAISocialSciences2019}'s warning that “qualitative social science disciplines like sociology and anthropology...are filtered out [while] quantitative approaches such as analytic philosophy, behavioural economics, and evolutionary psychology... dominate”, the paper suggests that the field of responsible AI development--understood as the building of AI systems which are safe, secure, and socially beneficial \citep{askellRoleCooperationResponsible2019}--might be poised to gain from engaged, embedded, qualitative work from within.

\section{Gaps in Practice}

Despite a growing list of tools and methods meant to aid the operationalisation of AI principles \citep{morleyWhatHowInitial2019}, not much is currently known about the extent of their implementation and its associated challenges. Some have argued that adoption is lagging behind due to high implementation costs; for example, \citet{guptaGreenLightingML2020} have suggested that the reason that documentation efforts \citep{hollandDatasetNutritionLabel2018,mitchellModelCardsModel2019,gebruDatasheetsDatasets2020} have not been implemented have been because they are “too onerous to create in practice”, particularly when the value of such production is not immediately evident. Indeed, the implementation of these tools often requires allocating, balancing, and trading off crucial resources--time and sometimes specific expertise--that individuals are hesitant to invest in. 

Nonetheless, while this gap may be explained by high demands placed on practitioners, more research might be done to understand these implementation frictions within organisations. That is to say, regardless of how well-intentioned, well-designed, or theoretically sound these tools and methods are; their successes or failures will not always be the result of decisions driven by individuals (be they researchers or practitioners), but rather the result of institutional and organisational norms and incentive structures. These norms and structures shape the openness, amenability, and willingness of teams to experiment with emerging practices that fit within current development pipelines.  

Already, some have highlighted the dangers of competitive market pressures that may disincentivise responsible practices \citep{askellRoleCooperationResponsible2019}. Indeed, one may expect that the reasons why documentation efforts are considered onerous may be attributed to tight timelines. Nonetheless, the interplay of new practices, individual interests, and competition pressures have often led to more complex and contingent practices instead. Within academic and industry laboratories, social relations, financial conditions, and technological affordances (both hardware and software) come together to shape the articulation and boundaries of normative and complex problems. A failure to come to terms with these differences may, in the long run, contribute further to the growing gap between the “ideals of deploying ethical, robust, and trustworthy ML and the practical reality of deploying ML systems” \citep{guptaGreenLightingML2020}. 

Doing so requires AI researchers to turn our gaze \textit{inwards}, towards the very relationships, norms, and structures that work both collectively and combatively to implement responsible AI practices. Social scientists are well-positioned to understand these very frictions. Indeed, within the field there has already been a turn in recent years towards using qualitative methods to engage with practitioners: \citet{vealeFairnessAccountabilityDesign2018} sought to reveal how public sector AI practitioners understood fairness and the challenges they faced in implementing fairness standards in their own work; \citet{holsteinImprovingFairnessMachine2019} similarly conducted research on the needs of practitioners within the private sector. Elsewhere, \citet{orrAttributionsEthicalResponsibility2020} have sought to identify how practitioners conceptualise accountability and responsibility within the AI systems that they work on; and more recently \citet{rakovaWhereResponsibleAI2020} conducted interviews in order to map organisational structures that currently support and hinder responsible AI initiatives. These studies have been illuminating in their findings, which have shown that practitioners are invested in the socio-political implications of their work and yet are often not best positioned to change or develop existing practices.

This inward, reflexive turn has already been preempted by scholars in the field: \citet[p.73]{elishSituatingMethodsMagic2018} previously invited practitioners to conceive of their work as a form of “computational ethnography” in order to “[explore and develop] what it means to be reflexive in the methods of data science and statistics”. \citet[p.208]{dignazioDataFeminism2020a} have similarly argued for more “deliberate interventions in each phase of data work, and in our received ideas about the people and communities who perform it”. These invitations, on the other hand, have to grapple with the reality of data science \textit{on the ground}: calls for more participatory, pluralist approaches to data science and AI research risk placing too much faith and expectations on individual practitioners who--despite being well-intentioned--often “lack the methodological tools necessary to critically engage with the epistemological and normative aspects of their work” \citep[p.170]{barabasStudyingReorientingStudy2020}. Once again, the ideals of deploying responsible AI practices bump up against the practical realities of who, when, and how these practices are adopted. 

One might then ask how existing research can go further. It is noteworthy that qualitative research on understanding responsible AI has concentrated on interviews with AI practitioners working across a variety of companies and industries; yet AI development and research is often a team endeavour. While the research above has been illustrative for understanding how practitioners conceptualise the scope of their jobs, these studies remain compartmentalised and insulated from practices that happen in day-to-day research. To better understand how ongoing collaborations and contestation yields responsible innovation, other forms of qualitative research--participant observation, document studies--can complement interviews to shed more light on how the community is coming to terms with the normative implications of their work. 

\section{Opening pathways for research}

Thus, potentially impactful and insightful research might be gleaned from ethnographic research conducted within AI research laboratories and organisations, such that implementation frictions can be further illuminated and understood. One might, for example, conceptualise the undertaking of “responsible AI” as a form of sensemaking \citep{weickSensemakingOrganizations1995}. While this paper has adopted the definition from \citet{askellRoleCooperationResponsible2019}'s work of responsible AI as the building of systems which are safe, secure, and socially beneficial; the term remains contested and open for debate. Disagreements over the trajectory of AI development continue to shape how researchers prioritise the very practices that fall under the broad umbrella of responsible AI \citep{prunklLongTermClearerAccount2020}. It is hypothesised that these disagreements may also shape the decisions that lead to the implementation (or non-implementation) of tools, practices, and communication policies that shape responsible innovation.    

This focus on sensemaking shifts the analytical gaze away from a needs-analysis \citep{vealeFairnessAccountabilityDesign2018,holsteinImprovingFairnessMachine2019} and away from the role of single individuals. Instead, it grounds existing practices within social, cultural, and institutional forms in order to bring to the fore asymmetries of power that open or close routes of action. Ethnographic work like this has a long history within science and technology studies (STS): from classic studies of human machine interactions \citep{orrTalkingMachinesEthnography1996,suchmanPlansSituatedActions1987} to more recent work dedicated to scrutinising the very processes of digitalisation \citep{grayGhostWorkHow2019, camus2019unfolding}. Researchers of emerging technologies at the turn of the millennium--nanotechnology \citep{johnsonEthicsTechnologyMaking2007} and the human genome project \citep{mcharekHumanGenomeDiversity2005}--have also grappled with questions of harm, risk, and impact that we ask of AI researchers. These questions have also been addressed in safety science--a field which has often used ethnographic methods to understand the successes and failures of organisations within high-risk industries (e.g., petroleum, chemicals, civil aviation, nuclear power plants) \citep{vaughanChallengerLaunchDecision1997,hopkinsDisastrousDecisionsHuman2012,haavikNewToolsOld2017} Despite a growing literature calling for more scrutiny of unspoken work practices of AI experts \citep{passiProblemFormulationFairness2019}, comparatively less work has been done so far dedicated towards examining the day-to-day practices of those involved at the forefront of AI development--be it within academia or industry. Instead, more work has been concentrated post-hoc, dedicated towards understanding the challenges of \textit{implementing} systems which have already been designed \citep{elishRepairingInnovationStudy,neffAIWorkArtificial}. 

Given this empirical gap, one way for the field to develop better insights on responsible AI research is to dedicate more resources towards projects that encourage "deep hanging out" further up the production pipeline \citep{geertzDeepHangingOut1998}. Doing so enables us to reframe current questions: not "\textit{what practices of responsible AI are available?}", but rather "\textit{how are practices of responsible AI produced?}" Efforts like these are already underway, \citet{dentonBringingPeopleBack2020} have embarked on a research project very much in line with this line of thinking by studying the genealogy of data typically applied to benchmark ML datasets. In doing so, they recast their critical gaze back onto those most involved in the production of AI/ML models, "creators of data collection, their taxonomic choices, decisions and their intended or unintended effects within a network of relations operative in a given data set”. 

This form of research has two contributions. Firstly, this builds further on existing literature calling for more user studies of existing tools \citep{gebruDatasheetsDatasets2020,kaurInterpretingInterpretabilityUnderstanding2020}. Whether through processual mechanisms or technical methods (e.g., glassbox models), the choice to adopt and adapt these tools is not itself an indicator that researchers have sufficiently engaged in the forms of reflexivity that scholars argue are essential for anticipating and addressing potential harms. As \citet{kaurInterpretingInterpretabilityUnderstanding2020} and \citet{dasOpportunitiesChallengesExplainable2020} have shown, the use of these tools may lead to cases of over-trust of the very systems meant to aid interpretability and explainability. Secondly, grounded research may also provide a more critical lens, enabling such work to sit within what Carly Kind has called the third wave of ethical AI: “beyond the principled and the technical” and focused on “questions of power, equity, and justice that underpin emerging technologies” \citep{kindTermEthicalAI2020}. More importantly, this shifts us away from focusing on individual psychologies--the form of analysis that dominates media coverage of questionable sociotechnical systems by over-emphasising a few bad eggs. Instead, it offers us an opportunity to focus on systemic norms and incentive structures are emerging within the community which "leave scientists and engineers vulnerable to ethical predicaments” \citep[p.308]{jeskeLessonsTheranosChanging2020}. 

\section{Challenges}

Recognising the potential contributions of ethnographic work also requires grappling with its challenges and limitations. Much more discussion is required than what can be contained here, but I offer the following: firstly, what forms of expertise are required? Researchers best positioned to conduct ethnographic research also require technical expertise and familiarity with day-to-day development and organisational practices. All of these factors vary across domains and sectors. This challenge may of course resolve itself if one remains embedded in a sector, organisation, or project team long enough to develop an “interactional expertise” \citep{doi:10.1177/0306312702032002003} with other players in the field. This raises a second challenge, which is one of access and reception. Will AI Research labs be open to such inquiry? As \citet{barabasStudyingReorientingStudy2020} have previously reflected in their attempt to “study up”, research that orients the analytical gaze upwards towards people and institutions of power often find themselves facing gatekeepers “who have the power to resist outside scrutiny”. Labs will have to assess sensitivity issues and whether they have sufficient resources to accommodate external researchers. It is expected that access to sites of research may face multiple structural barriers which are reflective of current asymmetries of power and resource distribution in the field (e.g., \citep{hooker2020hardware}). Given that the discussion of ethical principles has itself been dominated by a select number of institutions and countries \citep{jobinGlobalLandscapeAI2019}, this work may itself compound existing inequalities by crowding out less mainstream perspectives. This, in turn, would be counterproductive to the goal of showing how sociocultural variations deeply influence the trajectories of how knowledge, expertise, and practice interact to yield what ultimately becomes recognised as responsible practices in AI research and development. One solution here could be multi-sited comparisons across industries and across frequently underrepresented countries. Nonetheless, until this form of research is \textit{invited} in, researchers capable of this inquiry will be left with their noses pressed up against the glass of labs at the forefront of AI research.  

\section{Conclusion}
The move towards responsible AI has been increasingly gaining traction. While the development of new tools and practices meant to aid the realisation of responsible research is welcomed, knowledge of what it means to use these in practice remains relatively under-explored in the literature. On-site qualitative research--observations, interviews, document studies--can clarify this gap and enable the field to better understand the emerging discourse and practices of "responsible AI". More importantly, by illuminating the social nature of relationships, norms and structures that influence the trajectory of AI research, it is hoped that this form of inquiry can contribute to identifying different organisational and institutional challenges that the AI research community faces as it confronts its own role in addressing and mitigating algorithmic harms. As a field where innovations can potentially bring far-ranging consequences at scale, creating pathways for ethnographic work that spark organisational and institutional reflection could be one way of moving this conversation of responsible research to its next step.  

\newpage

\section*{Broader Impact}

This paper was written specifically to address the question of how to best move the discussion of responsible innovation forward so as to enable the AI research community to better anticipate and address potential harms of their research. Given the opaqueness that characterises AI/ML production and efforts towards transparency in the field, we hope that our suggestion of dedicating more resources (time, money, expertise) to on-site ethnographic work will enable greater clarity of the various frictions that undoubtedly mark critical reflections of research practices. It is hoped that the arguments in this paper will be relevant to both researchers and practitioners in industry, and may spark further discussions about how resources dedicated towards achieving this goal are not unevenly distributed among universities and commercial laboratories.   

\section*{Acknowledgements}
This research is supported by the National Research Foundation, Singapore under its Emerging Areas Research Projects (EARP) Funding Initiative. Any opinions, findings and conclusions or recommendations expressed in this material are those of the author and do not reflect the views of National Research Foundation, Singapore. 

Many thanks to the reviewers for their comments which helped improve the paper. I am also grateful to Mark Findlay, Alicia Wee, and Jane Loo for their invaluable feedback.

\bibliography{library.bib}

\end{document}